Big Data Lab
- Databases and Information Systems (DBIS) -

GOETHE UNIVERSITÄT FRANKFURT AM MAIN

# Performance Evaluation of Virtualized Hadoop Clusters

*Technical Report No. 2014-1*
*November 14, 2014*

Todor Ivanov, Roberto V. Zicari, Sead Izberovic,
Karsten Tolle

Frankfurt Big Data Laboratory
Chair for Databases and Information Systems
Institute for Informatics and Mathematics
Goethe University Frankfurt
Robert-Mayer-Str. 10,
60325 Bockenheim,
Frankfurt am Main, Germany
www.bigdata.uni-frankfurt.de

# Table of Contents

# 1. Introduction

Apache Hadoop [1] has emerged as the predominant platform for Big Data applications. Recognizing this potential, Cloud providers have rapidly adopted it as part of their services (IaaS, PaaS and SaaS)[2]. For example, Amazon, with its Elastic MapReduce (EMR) [3] web service, has been one of the pioneers in offering Hadoop-as-a-service. The main advantages of such cloud services are quick automated deployment and cost-effective management of Hadoop clusters, realized through the pay-per-use model. All these features are made possible by virtualization technology, which is a basic building block of the majority of public and private Cloud infrastructures [4]. However, the benefits of virtualization come at a price of an additional performance overhead. In the case of virtualized Hadoop clusters, the challenges are not only the storage of large data sets, but also the data transfer during processing. Related works, comparing the performance of a virtualized Hadoop cluster with a physical one, reported virtualization overhead ranging between 2-10% depending on the application type [5], [6], [7]. However, there were also cases where virtualized Hadoop performed better than the physical cluster, because of the better resource utilization achieved with virtualization.

In spite of the hypervisor overhead caused by Hadoop, there are multiple advantages of hosting Hadoop in a cloud environment [5], [6], [7] such as improved scalability, failure recovery, efficient resource utilization, multi-tenancy, security, to name a few. In addition, using a virtualization layer enables to separate the compute and storage layers of Hadoop on different virtual machines (VMs). Figure 1 depicts various combinations to deploy a Hadoop cluster on top of a hypervisor. Option (1) is hosting a worker node in a virtual machine running both a TaskTracker and NameNode service on a single host. Option (2) makes use of the multi-tenancy ability provided by the virtualization layer hosting two Hadoop worker nodes on the same physical server. Option (3) shows an example for functional separation of compute (MapReduce service) and storage (HDFS service) in separate VMs. In this case, the virtual cluster consists of two compute nodes and one storage node hosted on a single physical server. Finally, option (4) gives an example for two separate clusters running on different hosts. The first cluster consists of one data and one compute node. The second cluster consists of a compute node that accesses the data node of the first cluster. These deployment options are currently supported by Serengeti [8], a project initiated by VMWare, and Sahara [9], which is part of the OpenStack [10] cloud platform.

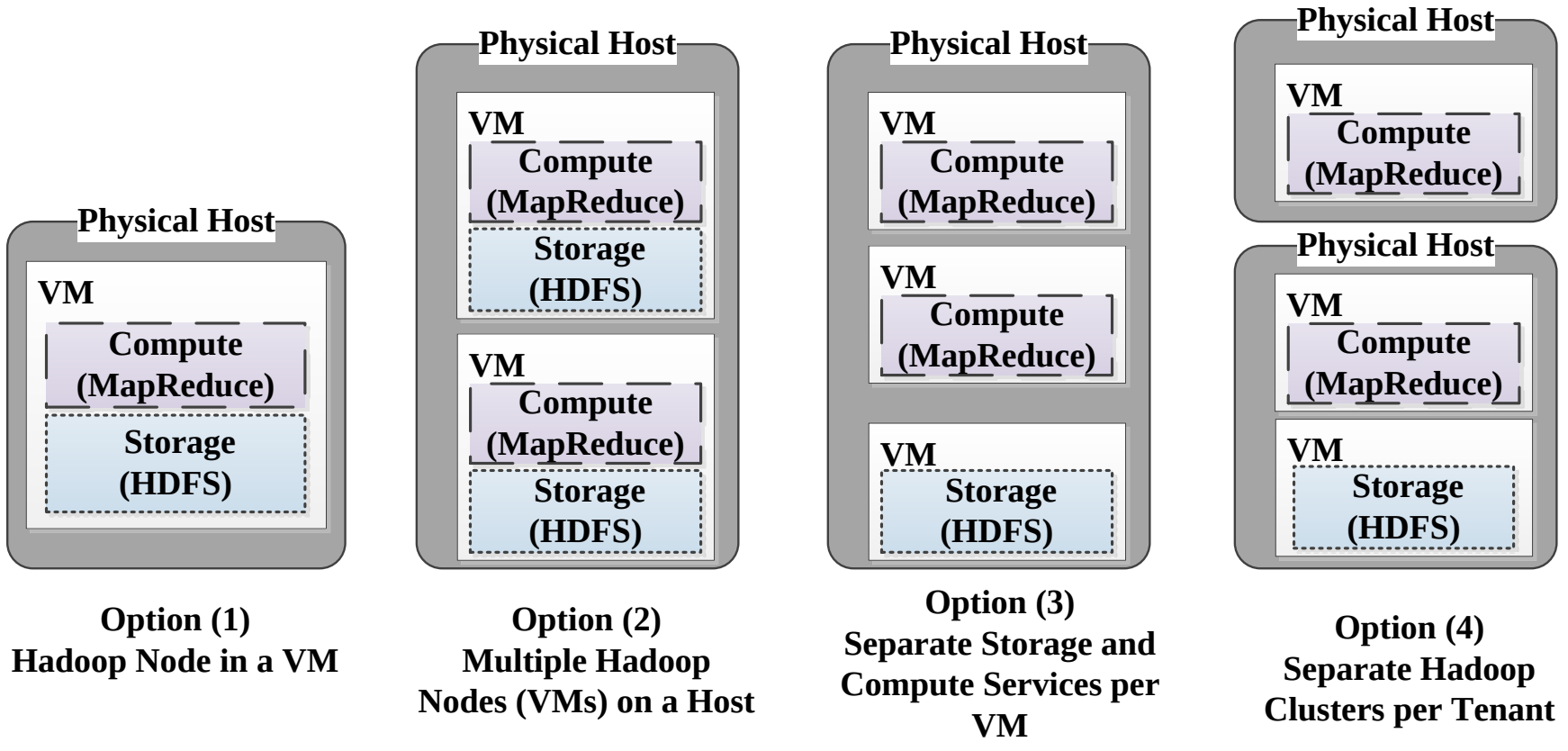

Figure 1: Options for Virtualized Hadoop Cluster Deployments

In this report we investigate the performance of Hadoop clusters, deployed with separated storage and compute layers (option (3)), on top of a hypervisor managing a single physical host. We have analyzed and evaluated the different Hadoop cluster configurations by running CPU bound and I/O bound workloads.
The report is structured as follows: Section 2 provides a brief description of the technologies involved in our study. An overview of the experimental platform, setup test and configurations are presented in Section 3. Our benchmark methodology is defined in Section 4. The performed experiments together with the evaluation of the results are presented in Section 5. Finally, Section 6 concludes with lessons learned.

## 2. Background

**Big Data** has emerged as a new term not only in IT, but also in numerous other industries such as healthcare, manufacturing, transportation, retail and public sector administration [11], [12] where it quickly became relevant. There is still no single definition which adequately describes all Big Data aspects [13], but the "*V*" characteristics (*Volume*, *Variety*, *Velocity*, *Veracity* and more) are among the widely used one. Exactly these new Big Data characteristics challenge the capabilities of the traditional data management and analytical systems [13], [14]. These challenges also motivate the researchers and industry to develop new types of systems such as Hadoop and NoSQL databases [15].

**Apache Hadoop** [1] is a software framework for distributed storing and processing of large data sets across clusters of computers using the map and reduce programming model. The architecture allows scaling up from a single server to thousands of machines. At the same time Hadoop delivers high-availability by detecting and handling failures at the application layer. The use of data replication guarantees the data reliability and fast access. The core Hadoop components are the Hadoop Distributed File System (HDFS) [16], [17] and the MapReduce framework [18].
HDFS has master/slave architecture with a *NameNode* as a master and multiple *DataNodes* as slaves. The *NameNode* is responsible for the storing and managing of all file structures, metadata, transactional operations and logs of the file system. The *DataNodes* store the actual data in the form of files. Each file is split into blocks of a preconfigured size. Every block is copied and stored on multiple *DataNodes*. The number of block copies depends on the *Replication Factor*.
MapReduce is a software framework, that provides general programming interfaces for writing applications that process vast amounts of data in parallel, using a distributed file system, running on the cluster nodes. The MapReduce unit of work is called *job* and consists of input data and a MapReduce program. Each job is divided into *map* and *reduce* tasks. The map task takes a split, which is a part of the input data, and processes it according to the user-defined map function from the MapReduce program. The reduce task gathers the output data of the map tasks and merges them according to the user-defined reduce function. The number of reducers is specified by the user and does not depend on input splits or number of map tasks. The parallel application execution is achieved by running map tasks on each node to process the local data and then send the result to a reduce task which produces the final output.
Hadoop implements the MapReduce model by using two types of processes – *JobTracker* and *TaskTracker*. The *JobTracker* coordinates all jobs in Hadoop and schedules tasks to the *TaskTrackers* on every cluster node. The *TaskTracker* runs tasks assigned by the *JobTracker*.
Multiple other applications were developed on top of the Hadoop core components, also known as the Hadoop ecosystem, to make it more ease to use and applicable to variety of industries. Ex-ample for such applications are Hive [19], Pig [20], Mahout [21], HBase [22], Sqoop [23] and many more.

**VMware vSphere** [24]**,** [25] is the leading server virtualization technology for cloud infrastructure, which consisting of multiple software components with compute, network, storage, availability, automation, management and security capabilities. It virtualizes and aggregates the underlying physical hardware resources across multiple systems and provides pools of virtual resources to the datacenter.

**Serengeti** [8] is an open source project started by VMware and now part of the vSphere Big Data Extension [26]. The goal of the project is to enable quick configuration and automated deployment of Hadoop in virtualized environments. The major contribution of the project is the Hadoop Virtual Extension (HVE) [27], which makes Hadoop aware that it is virtualized. This new layer integrating hypervisor functionality is implemented using hooks that touch all of the Hadoop subcomponents (Common, HDFS and MapReduce) and is called *Node Group layer*. Additionally, new data-locality related policies are included: ***replica placement*** /***removal*** policy extension, ***replica choosing*** policy extension and ***balancer*** policy extension. According to the VMware report [28], the benefits of virtualizing Hadoop are: (i) enabling rapid provisioning;(ii) additional high availability and fault tolerance provided by the hypervisor;(iii) improving datacenter efficiency by higher server consolidation;(iv) efficient resource utilization by guaranteeing virtual machines resources;(v) multi-tenancy allowing mixed workloads on the same tenant but still preserving the Quality of Service (QoS) and SLA's; vi) provides security and isolation between the virtual machines;(vii) enables time sharing by scheduling jobs to run in periods with low hardware usage;(viii) easy maintenance and movement of environment;(ix) enables to run Hadoop-as-a-service in Cloud environment. Another major functionality that Serengeti introduces for the first time is the ability to separate the compute and storage layers of Hadoop on different virtual machines.

## 3. Experimental Environment

### 3.1. Platform

An abstract view of the experimental platform we used to perform the tests is shown in Figure 2. The platform is organized in four logical layers which are described below.

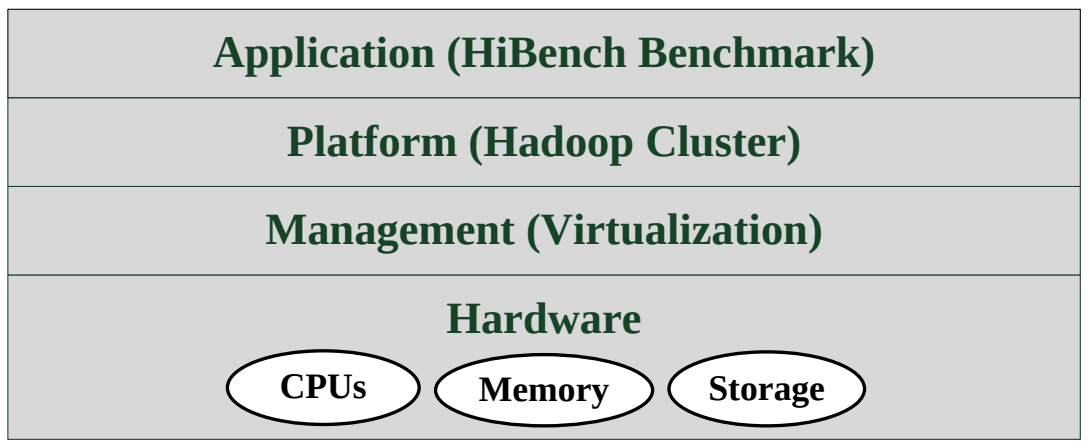

Figure 2: Experimental Platform Layers

**Hardware**
It consists of a standard Dell PowerEdge T420 server equipped with two Intel Xeon E5-2420 (1.9 GHz) CPUs each with six cores, 32 GB of RAM and four 1 TB, Western Digital (SATA, 3.5 in, 7.2K RPM, 64MB Cache) hard drives.

**Management (Virtualization)**
We installed the VMware vSphere 5.1 [24] platform on the physical server, including ESXi and vCenter Servers for automated VM management.

**Platform (Hadoop Cluster)**
Project Serengeti integrated in the vSphere Big Data Extension (BDE) (version 1.0) [26], installed in a separate VM, was used for automatic deployment and management of Hadoop clusters. The hard drives were deployed as separate data stores and used as shared storage resources by BDE. The deployment of both Standard and Data-Compute cluster configurations was done using the default BDE/Serengeti Server options as described in [29]. In all the experiments we used the Apache Hadoop distribution (version 1.2.1), included in the Serengeti Server VM template (hosting CentOS), with the default parameters: 200MB java heap size, 64MB HDFS block size and *Replication Factor* of 3.

**Application (HiBench Benchmark)**
The HiBench [30] benchmark suite was develop by Intel to stress test Hadoop systems. It contains 10 different workloads divided in 4 categories:

1. Micro Benchmarks (Sort, WordCount, TeraSort, Enhanced DFSIO)
2. Web Search (Nutch Indexing, PageRank)
3. Machine Learning (Bayesian Classification, K-means Clustering)
4. Analytical Queries (Hive Join, Hive Aggregation)

For our experiments, we have chosen two MapReduce representative applications from the HiBench micro-benchmarks, namely, the *WordCount* (CPU bound) and the *TestDFSIOEnhanced* (I/O bound) workloads.
One obvious limitation of our experimental environment is that it consists of a single physical server, hosting all VMs, and does not involve any physical network communication between the VM nodes. Additionally, all experiments were performed on the VMware ESXi hypervisor. This means that the reported results may not apply to other hypervisors as suggested by related work [31], comparing different hypervisors.

## 3.2. Setup and Configuration

The focus of this report is on analyzing the performance of different virtualized Hadoop cluster configurations, deployed and tested on our platform. Figure 3 shows the two types of cluster configurations investigated in this report, namely: *Standard* and *Data-Compute* clusters.

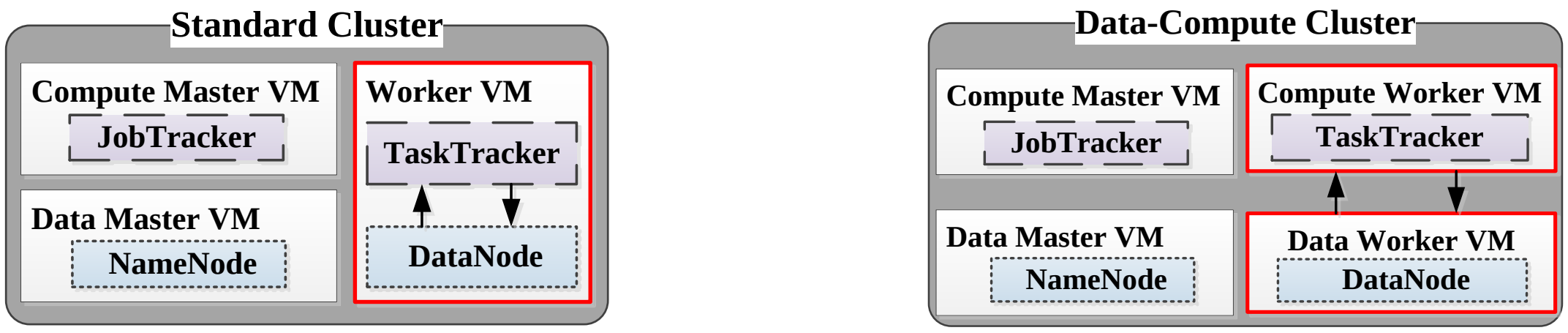

Figure 3: Standard and Data-Compute Hadoop Cluster Configurations

The *Standard Hadoop* cluster type is a standard Hadoop cluster configuration but hosted in a virtualized environment with each cluster node installed in a separate VM. The cluster consists of one Compute Master VM (running *JobTracker*), one Data Master VM (running *NameNode*) and multiple Worker VMs. Each Worker VM is running both *TaskTracker* and *DataNode* services. *The data exchange is between the TaskTracker and DataNode services inside the VM.*

On the other hand, the *Data-Compute Hadoop* cluster type has similarly Compute and Data Master VMs, but two types of Worker nodes: Compute and Data Worker VMs. This means that there are data nodes, running only *DataNode* service and compute nodes, running only *TaskTracker* service. *The data exchange is between the Compute and Data VMs, incurring extra virtual network traffic.* The advantage of this configuration is that the number of data and compute nodes in a cluster can be independently and dynamically scaled, adapting to the workload requirements.

The first factor that we have to take into account when comparing the configurations is *the number of VMs utilized in a cluster*. Each additional VM increases the hypervisor overhead and therefore can influence the performance of a particular application as reported in [31], [6], [32]. At the same time, running more VMs utilizes more efficiently the hardware resources and in many cases leads to improved overall system performance (CPU and I/O Throughput) [6].

The second factor is that all cluster configurations should *utilize the same amount of hardware resources* in order to be comparable.

Taking these two factors into account, we specified six different cluster configurations. Two of the cluster configurations are of type *Standard Hadoop* cluster and the other four are of type *Data-Compute Hadoop* cluster. Based on the number of virtual nodes utilized in a cluster configuration, we compare *Standard1 with Data-Comp1* and *Standard2 with Data-Comp3* and *Data-Comp4*. Additionally, we added *Data-Comp2* to compare it with *Data-Comp1* and *Data-Comp3*. The goal is to better understand how the number of data nodes influences the performance of I/O bound applications in a *Data-Compute Hadoop* cluster.

Table 1 shows the worker nodes for each configuration and the allocated per VM resources (vCPUs, vRAM and vDisks). Three additional VMs (Compute Master, Data Master and Client VMs), not listed in Table 1, were used in all of the six cluster configurations. The exact parameters of each cluster configuration are described in a JSON file, which ensures the repeatability of the configuration resources and options. As an example the JSON file of the Data-Comp1 cluster configuration is included in the Appendix. For simplicity, we will abbreviate in the rest, the *Worker Node as WN*, the *Compute Worker Node as CWN* and *Data Worker Node as DWN*.

<table>
<tr><th>Configuration Name</th><th colspan="2">Worker Nodes</th></tr>
<tr><td rowspan="2">Standard1<br>(Standard Cluster 1)</td><td colspan="2">3 Worker Nodes</td></tr>
<tr><td colspan="2">TaskTracker & DataNode; 4 vCPUs; 4608MB vRAM; 100GB vDisk</td></tr>
<tr><td rowspan="2">Standard2<br>(Standard Cluster 2)</td><td colspan="2">6 Worker Nodes</td></tr>
<tr><td colspan="2">TaskTracker & DataNode; 2 vCPUs; 2304MB vRAM; 50GB vDisk</td></tr>
<tr><td rowspan="2">Data-Comp1<br>(Data-Compute Cluster 1)</td><td>2 Compute Worker Nodes</td><td>1 Data Worker Node</td></tr>
<tr><td>TaskTracker; 5 vCPUs;<br>4608MB vRAM; 50GB vDisk</td><td>DataNode; 2 vCPUs; 4608MB vRAM;<br>200GB vDisk</td></tr>
<tr><td rowspan="2">Data-Comp2<br>(Data-Compute Cluster 2)</td><td>2 Compute Worker Nodes</td><td>2 Data Worker Nodes</td></tr>
<tr><td>TaskTracker; 5 vCPUs;<br>4608MB vRAM; 50GB vDisk</td><td>DataNode; 1 vCPUs;<br>2084MB vRAM; 100GB vDisk</td></tr>
<tr><td rowspan="2">Data-Comp3<br>(Data-Compute Cluster 3)</td><td>3 Compute Worker Nodes</td><td>3 Data Worker Nodes</td></tr>
<tr><td>TaskTracker; 3 vCPUs;<br>2664MB vRAM; 20GB vDisk</td><td>DataNode; 1 vCPUs;<br>1948MB vRAM; 80GB vDisk</td></tr>
<tr><td rowspan="2">Data-Comp4<br>(Data-Compute Cluster 4)</td><td>5 Compute Worker Nodes</td><td>1 Data Worker Nodes</td></tr>
<tr><td>TaskTracker; 2 vCPUs;<br>2348MB vRAM; 20GB vDisk</td><td>DataNode; 2 vCPUs;<br>2048MB vRAM; 200GB vDisk</td></tr>
</table>

Table 1: Six Experimental Hadoop Cluster Configurations

## 4. Benchmarking Methodology

In this section we describe our benchmarking methodology that we defined and used throughout all experiments. The major motivation behind it was to ensure the comparability between the measured results.

We started by selecting 2 out of the 10 HiBench [30] workloads as listed in Table 2. Our goal was to have representative workloads for CPU bound and I/O bound workloads.

| Workload | Data structure | CPU usage | IO (read) | IO (write) |
|---|---|---|---|---|
| WordCount | unstructured | high | low | low |
| Enhanced DFSIO | unstructured | low | high | high |

Table 2: Selected HiBench Workload Characteristics

Figure 4 briefly illustrates the five phases in our experimental methodology, which we call an *Iterative Experimental Approach*. In the initial Phase 1, all software components (VMware vSphere, Big Data Extension and Serengeti Server) are installed and configured.

In Phase 2, we setup the Apache Hadoop cluster using the Big Data Extension and Serengeti Server. We choose the cluster type, configure the number of nodes and set the virtualized resources as listed in Table 1. Finally, the cluster configuration is created and HiBench is installed in the client VM.

Next in Phase 3, called Workload Prepare, are defined all workload parameters and is generated the test data. The generated data together with the defined parameters are then used as input to execute the workload in Phase 4. As already mentioned each experiment was repeated 3 times to ensure the representativeness of the results, which means that the data generation from Phase 2 and the Workload Execution (Phase 4) were run 3 consecutive times. Before each workload experiment in the Workload Prepare (Phase 3), the existing data is deleted and new one is generated.

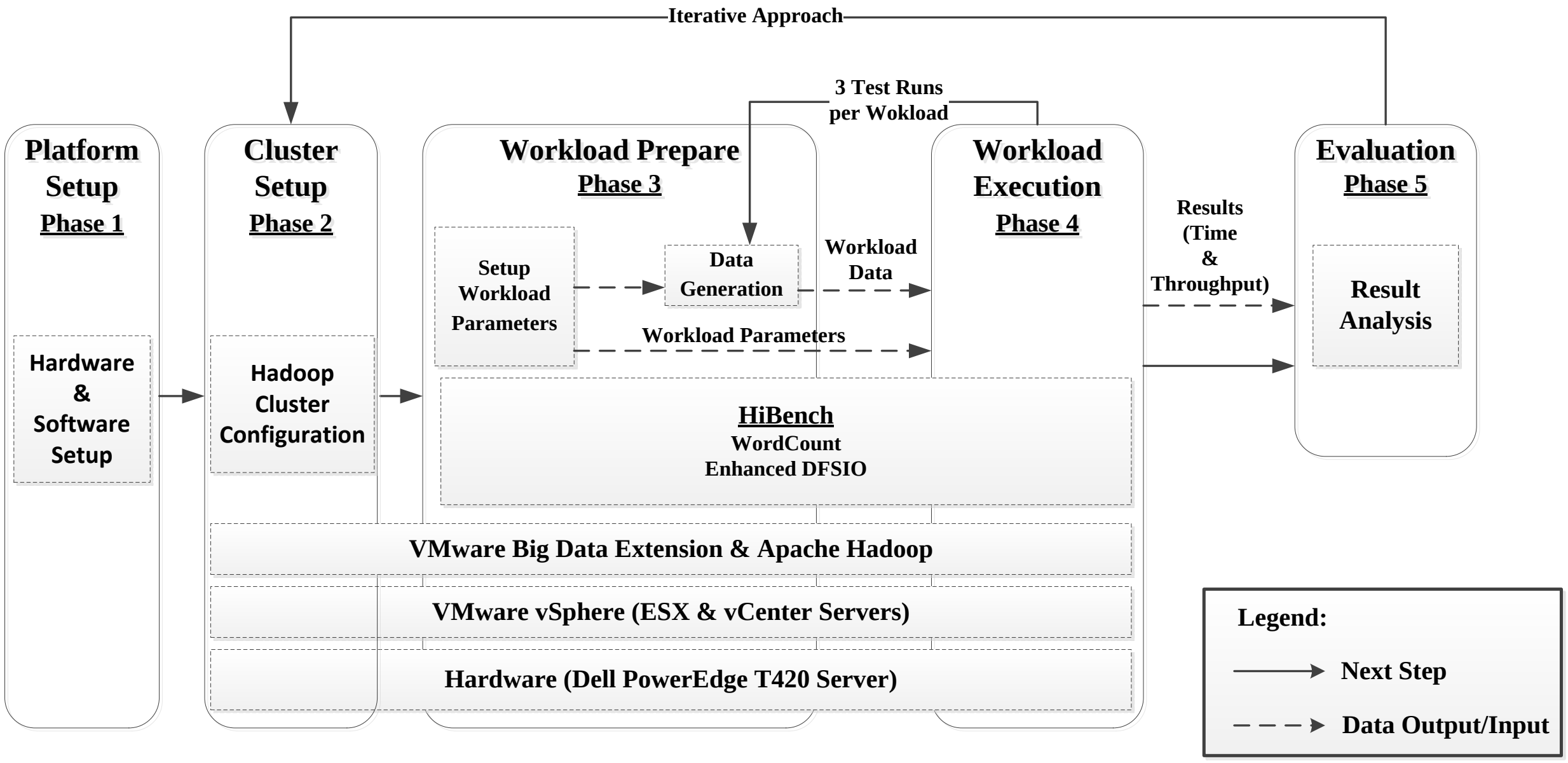

Figure 4: Iterative Experimental Approach

In Phase 4, HiBench reports two types of results: *Duration* (in seconds) and *Throughput* (in MB per second). The *Throughput* is calculated by dividing the input data size through the *Duration*. These results are then analyzed in Phase 5, called Evaluation, and presented graphically in the next section of our report.

We call our approach iterative because after the each test run (completing Phases 2 to 5) the user can start at Phase 2, switching to a different HiBench workload, and continue performing new test runs on the same cluster configuration. However, to ensure consistent data state, a fresh copy of the input data has to be generated before each benchmark run. Similarly, in case of new cluster configuration all existing virtual nodes have to be deleted and replaced with new ones, using a basic virtual machine template. For all experiments, we ran exclusively only one cluster configuration at a time on the platform. In this way, we avoided biased results due to inconsistent system state.

## 5. Experimental Results

This section gives a brief overview of the WordCount and Enhanced DFSIO workloads. It also presents the results and analysis of the performed experiments. The results are provided in tables, which consist of multiple columns with the following data:

- **Data Size (GB):** size of the input data in gigabytes
- **Time (Sec):** workload execution duration time in seconds
- **Data Δ (%):** difference of **Data Size (GB)** to a given data baseline in percent
- **Time Δ (%):** difference of **Time (Sec)** to a given time baseline in percent

## 5.1. WordCount

**WordCount** [30] is a CPU bound MapReduce job which calculates the number of occurrences of each word in a text file. The input text data is generated by the RandomTextWriter program which is also part of the standard Hadoop distributions.

### 5.1.1. Preparation

The workload takes three parameters listed in Table 3. The DATASIZE parameter is relevant only for the data generation.

| **Parameter** | **Description** |
|---|---|
| NUM_MAPS | Number of map jobs per node |
| NUM_REDS | Number of reduce jobs per node |
| Relevant for the data generator | |
| DATASIZE | Input of (text) data size per node |

Table 3: WordCount Parameters

In the case of a Data-Compute cluster, these parameters are only relevant for the Compute Workers (CWN running *TaskTracker*). Therefore, in order to achieve comparable results between the Standard and Data-Compute Hadoop cluster types, the *overall sum of the processed data and number of map and reduce tasks should be the same*. The total data size is equal to the DATASIZE multiplied by the number of CWN or WN. For example, to process 60GB data in Standard1 (3 WNs) cluster were configured 20GB input data size, 4 map and 1 reduce tasks, whereas in Data-Comp1 (2 CWNs & 1 DWN) cluster were configured 30GB input data size, 6 map and 1 reduce tasks. Similarly, we adjusted the input parameters for the remaining four clusters to ensure that the same amount of data was processed. We experimented with three different data sets (60, 120 and 180 GB), which compressed resulted in smaller sets (15.35, 30.7 and 46 GB).

### 5.1.2. Results and Evaluation

The following subsections represent different viewpoints of the same experiments and hence are based on same numbers. In the first subsection we compare the performance between the Standard and Data-Computer cluster configurations. The second subsection evaluates how increasing the data size changes the performance for each cluster configuration.

#### 5.1.2.1. Comparing Different Cluster Configurations

Figure 5 depicts the WordCount completion times normalized for each input data size with respect to Standard1 as baseline. *The lower values represent faster completion times, respectively the higher values account for longer completion times.*

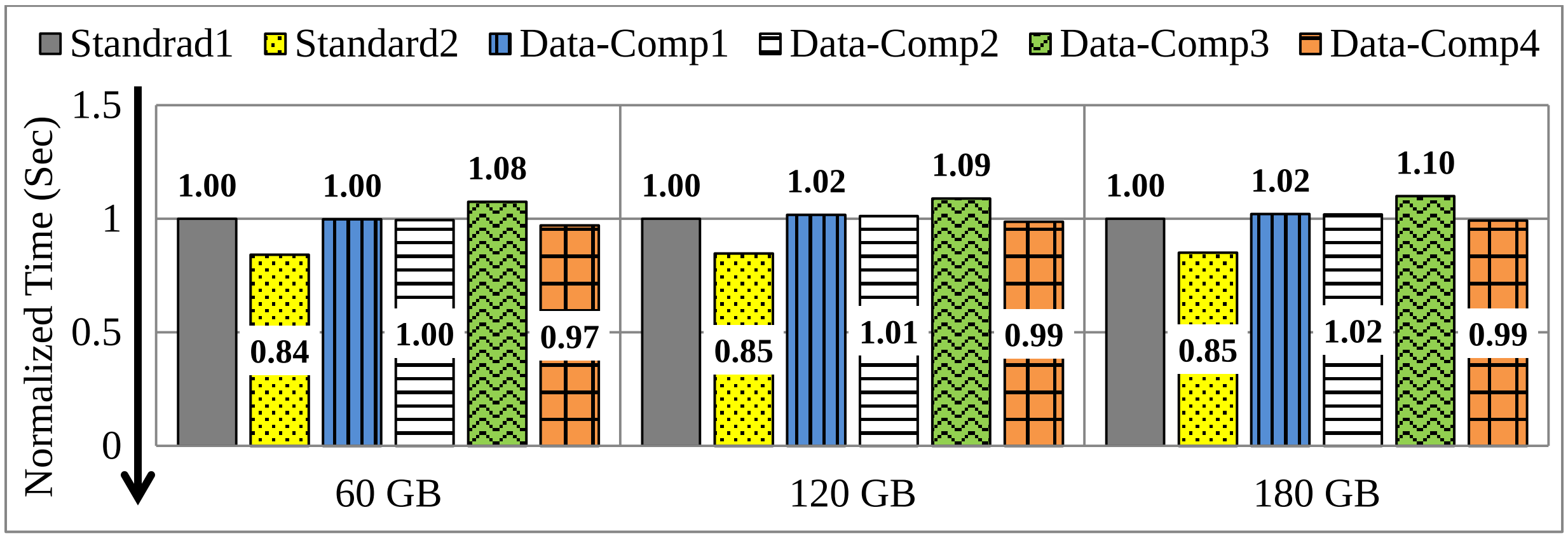

Figure 5: Normalized WordCount Completion Times

| Equal Number of VMs | 3 VMs | 6 VMs | 6 VMs |
|---|---|---|---|
| **Data Size (GB)** | **Diff. (%) Standard1/ Data-Comp1** | **Diff. (%) Standard2/ Data-Comp3** | **Diff. (%) Standard2/ Data-Comp4** |
| 60 | 0 | +22 | +13 |
| 120 | +2 | +22 | +14 |
| 180 | +2 | +23 | +14 |

Table 4. WordCount - *Equal Number of VMs*

| Different Number of VMs | 3 VMs / 6 VMs | 3 VMs / 6 VMs |
|---|---|---|
| **Data Size (GB)** | **Diff. (%) Standard1/ Standard2** | **Diff. (%) Standard1/ Data-Comp4** |
| 60 | -19 | -3 |
| 120 | -18 | -1 |
| 180 | -17 | -1 |

Table 5. WordCount - *Different Number of VMs*

Table 4 compares cluster configurations utilizing the same number of VMs. In the first case, Standard1 (3 WNs) performs slightly (2%) better than Data-Comp1 (2 CWNs & 1 DWN). In the second and third case, Standard2 (6 WNs) is around 23% faster than Data-Comp3 (3 CWNs & 3 DWNs) and around 14% faster than Data-Comp4 (5 CWNs & 1 DWNs), making it the best choice for CPU bound applications.

In Table 5, comparing the configurations with different number of VMs, we observe that Standard2 (6 WNs) is between 17-19% faster than Standard1 (3 WNs), although Standard2 utilizes 6 VMs and Standard1 only 3 VMs. Similarly, Data-Comp4 (5 CWNs & 1 DWNs) achieves between 1-3% faster times than Standard1 (3 WNs). In both cases having more VMs utilizes better the underlying hardware resources, which complies to the conclusions reported in [6].

Another interesting observation, as seen in Figure 5, is that cluster Data-Comp1 (2 CWNs & 1 DWN) and Data-Comp2 (2 CWNs & 2 DWN) perform alike, although Data-Comp2 utilizes an additional instance of data worker node, which causes extra overhead on the hypervisor. However, as the *WordCount* workload is mostly CPU bound [30], all the processing is performed on the compute worker nodes and the extra VM instance does not impact the actual performance. In the same time, if we compare the times on Figure 5 of all four Data-Compute cluster configurations, we observe that the Data-Comp4 (5 CWNs & 1 DWNs) performs best. This shows first that the allocation of virtualized resources influence the application performance and second that for CPU bound applications having more compute nodes is beneficial.

Serengeti offers the ability for Compute Workers to use a Network File System (NFS) instead of virtual disk storage, also called TempFS in Serengeti. The goal is to ensure data locality, increase capacity and flexibility with minimal overhead. A detailed evaluation and experimental results of the approach are presented in the related work [33]. Using the TempFS storage type, we performed experiments with Data-Comp1 and Data-Comp4 cluster configurations. The results showed very slight around 1% improvement compared to the default shared virtual disk type that we used in all configurations.

#### 5.1.2.2. Processing Different Data Sizes

Figure 6 depicts the *WordCount* processing times (in seconds) for the different data sizes of all the six cluster configurations. *The shorter times indicate better performance and respectively the longer times indicate worse performance.* Clearly, cluster configuration Standard2 (6 WNs) achieves the fastest times for all the three data sizes compared to the other configurations. This is also observed on Figure 7, which illustrates the throughputs (MBs per second) of all six configurations, where configuration Standard2 (6 WNs) achieves the highest throughput between 52-54 MBs per second.

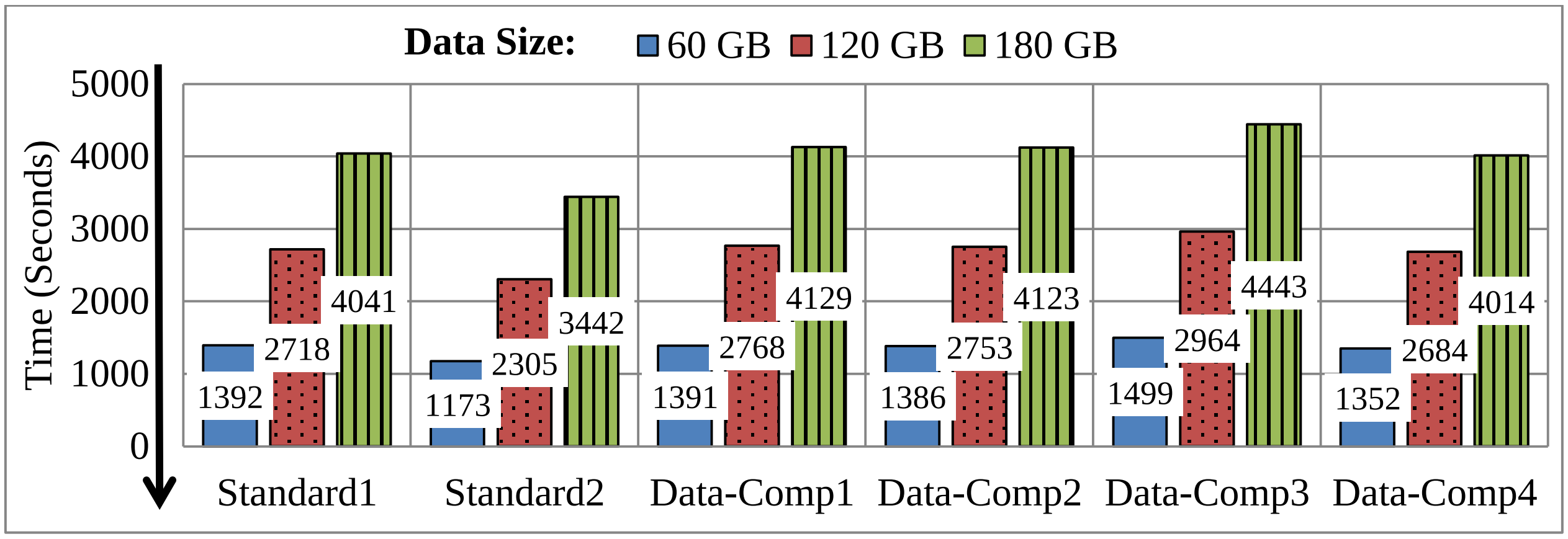

Figure 6: WordCount Time (Seconds)

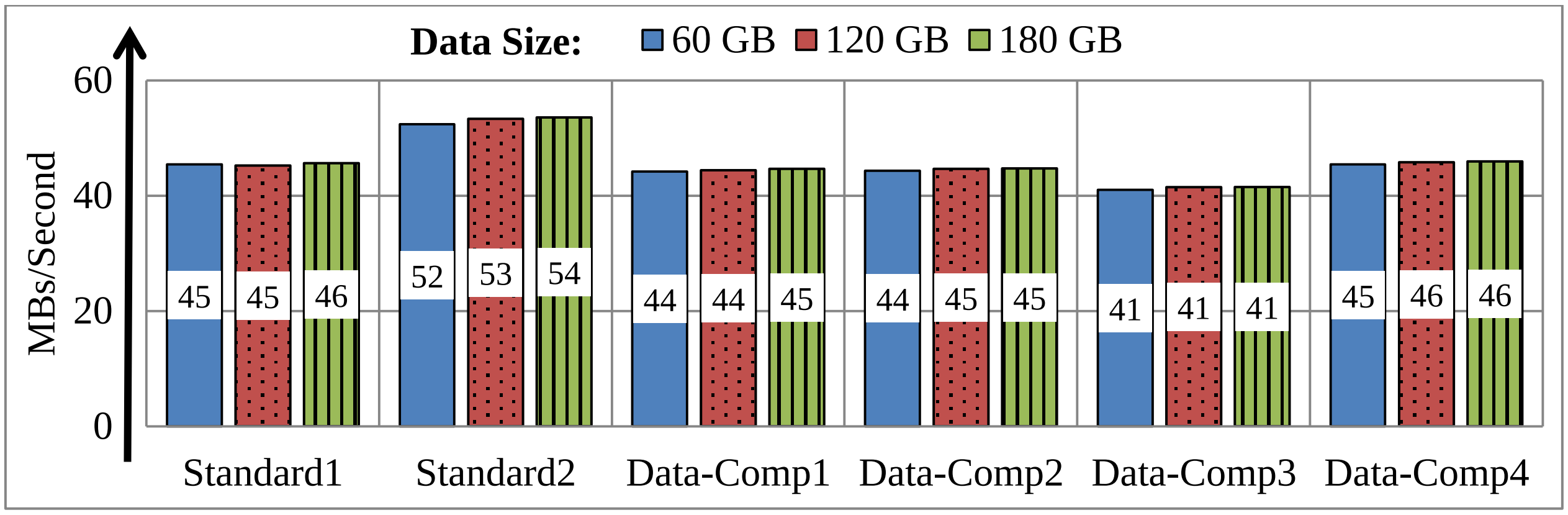

Figure 7: WordCount Throughput (MBs per second)

Table 6 and Table 7 summarize the processing times for all Standard and Data-Compute cluster configurations. Additionally, there is a column "**Data Δ**" representing the data increase in percent compared to the baseline data size, which is 60GB. For example, the **Δ** between the baseline (60GB) and 120GB is +100% and respectively for 180GB is +200%. Also, there are multiple columns "**Time Δ**", one per cluster configuration, which indicates the time difference in percent compared to the time of Standard1 (3 WNs), which we use as the *baseline configuration*. For example, comparing the times for processing 60GB of the baseline Standard1 (3 WNs) with Standard2 (6 WNs) configuration results in -15.72% time difference. This means that Standard2 finish-es for 15.72% less time compared to the baseline Standard1. Similarly, the positive time differences indicate slower completion times in comparison to the baseline.

| Data Size (GB) | Data Δ (%) | Standard1 (Sec) Baseline | Standard2 (Sec) | Time Δ (%) |
|---|---|---|---|---|
| 60 | baseline | 1392.06 | 1173.29 | -15.72 |
| 120 | +100 | 2718.03 | 2304.94 | -15.20 |
| 180 | +200 | 4040.5 | 3442.03 | -14.81 |

Table 6: WordCount Standard Cluster Results

| Data Size (GB) | Data Δ (%) | Data-Comp1 (Sec) | Time Δ (%) | Data-Comp2 (Sec) | Time Δ (%) | Data-Comp3 (Sec) | Time Δ (%) | Data-Comp4 (Sec) | Time Δ (%) |
|---|---|---|---|---|---|---|---|---|---|
| 60 | baseline | 1390.74 | -0.09 | 1385.63 | -0.46 | 1497.53 | +7.58 | 1351.87 | -2.89 |
| 120 | +100 | 2767.91 | +1.84 | 2752.92 | +1.28 | 2963.72 | +9.04 | 2684.3 | -1.24 |
| 180 | +200 | 4125.48 | +2.10 | 4122.59 | +2.03 | 4443.24 | +9.97 | 4013.77 | -0.66 |

Table 7: WordCount Data-Compute Cluster Results

Figure 8 depicts the time differences in percent of all cluster configurations normalized with respect to Standard1 (3 WNs) as baseline. We observe that Standard2 (6 WNs) and Data-Comp4 (5 CWNs & 1 DWNs) have *negative time difference*, which means that they perform faster than Standard1. On the other hand, Data-Comp1 (2 CWNs & 1 DWN), Data-Comp2 (2 CWNs & 2 DWN) and Data-Comp3 (3 CWNs & 3 DWNs) have *positive time differences*, which mean that they perform slower than Standard1.

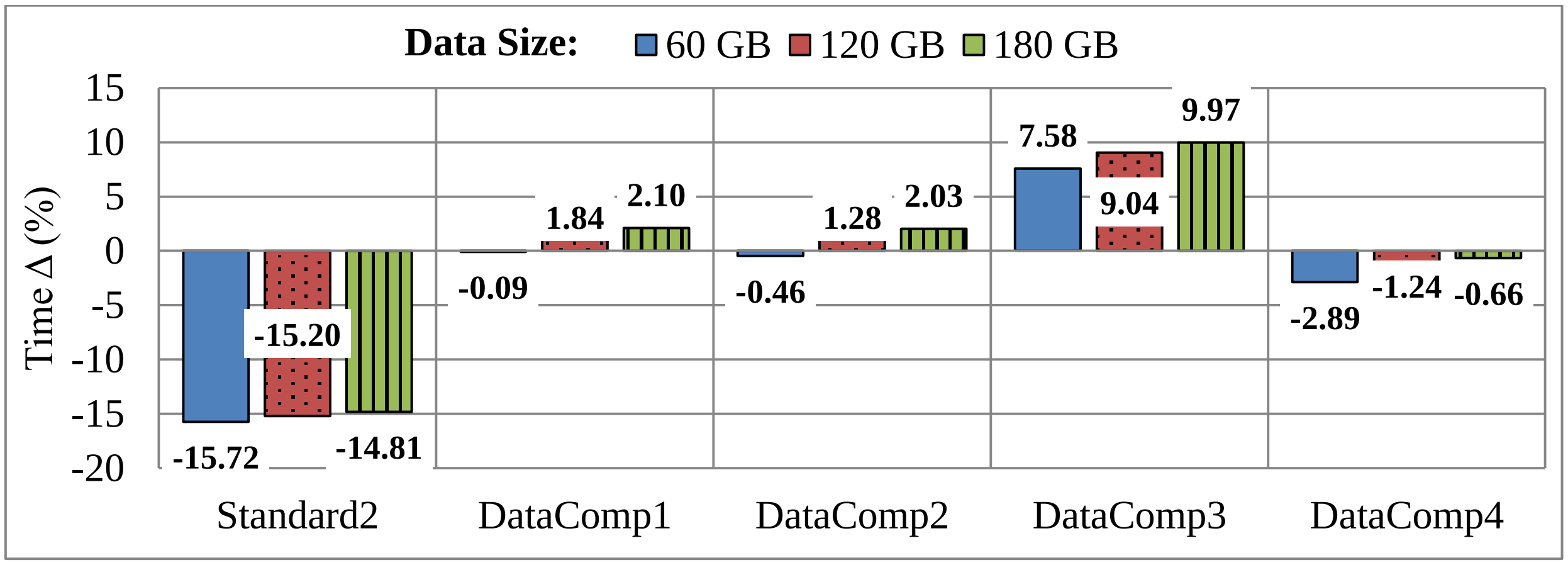

Figure 8: WordCount Time Difference between Standard1 (Baseline) and all Other Configurations in %

Figure 9 illustrates how the different cluster configurations scale with the increasing data sets normalized to the baseline configuration. We observe that all configurations scale nearly linear with the increase of the data sizes. However, similar to Figure 8 we can clearly distinguish that Standard2 (6 WNs) is the fastest configuration as its data points lie much lower than the other configurations. On the contrary, Data-Comp3 (3 CWNs & 3 DWNs) is the slowest configuration as its data points are the highest one for all three data sizes.

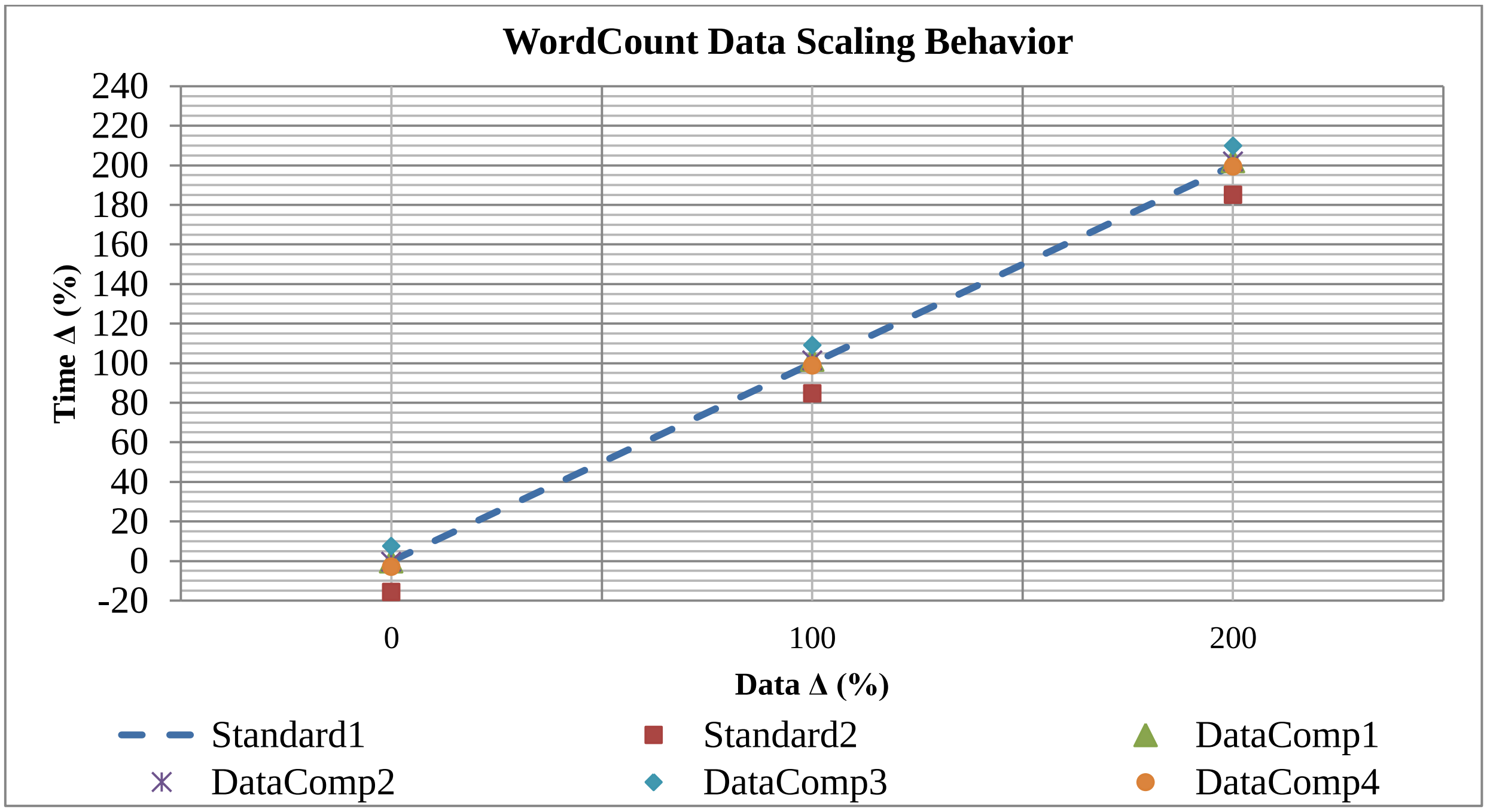

Figure 9: WordCount Data Scaling Behavior of all Cluster Configurations normalized to Standrad1

In summary, our experiments showed that for all cluster configurations the CPU bound WordCount workload scales nearly linear with the increase of the input data sets. Also we clearly observed that the Standard2 (6 WNs) configuration performs best, achieving the fastest times, whereas the Data-Comp3 (3 CWNs & 3 DWNs) performs worst, achieving the slowest completion time.

## 5.2. Enhanced DFSIO

**TestDFSIO** [34] is a HDFS benchmark included in Hadoop distributions. It is designed to stress test the storage I/O (read and write) capabilities of a cluster. In this way performance bottlenecks in the network, hardware, OS or Hadoop setup can be found and fixed. The benchmark consists of two parts: TestDFSIO-write and TestDFSIO-read. The write program starts multiple map tasks with each task writing a separate file in HDFS. The read program starts multiple map tasks with each task sequentially reading the previously written files and measuring the ***file size*** and the task ***execution time***. The benchmark uses a single reduce task to measure and compute two performance metrics for each map task: ***Average I/O Rate*** and ***Throughput***. Respectively, Equation 1 and Equation 2 illustrate how the two metrics are calculated with **N** as the total number of map tasks and the index **i** ($0 < i < N$), identifying the individual tasks.

Equation 1: ***Average I/O Rate***

$$\textbf{\textit{Average I/O rate (N)}} = \frac{\sum_{i=1}^{N} rate(i)}{N} = \frac{\sum_{i=1}^{N} \frac{file\ size(i)}{time(i)}}{N}$$

Equation 2: ***Throughput***

$$\textbf{\textit{Throughput (N)}} = \frac{\sum_{i=1}^{N} file\ size(i)}{\sum_{i=1}^{N} time(i)}$$

**Enhanced DFSIO** is an extension of the DFSIO benchmark developed specifically for HiBench [30]. The original TestDFSIO benchmark reports the average I/O rate and throughput for a single map task, which is not representative in cases when there are delayed or re-tried map tasks. Enhanced DFSIO addresses the problem by computing the aggregated I/O bandwidth. This is done by sampling the number of bytes read/written at fixed time intervals in the format of (map id, timestamp, total bytes read/written). Aggregating all sample points for each map tasks allows plotting the exact map task throughput as linearly interpolated curve. The curve consists of a warm-up phase and a cool-down phase, where the map tasks are started and shut down, respectively. In between is the steady phase, which is defined by a specified percentage (default is 50%, but can be configured) of map tasks. When the number of concurrent map tasks at a time slot is above the specified percentage, the slot is considered to be in the steady phase. The Enhanced DFSIO aggregated throughput metric is calculated by averaging value of each time slot in the steady phase.

### 5.2.1. Preparation

The Enhanced DFSIO takes four input configuration parameters as described in Table 8.

| Parameter | Description |
|---|---|
| RD_FILE_SIZE | Size of a file to read in MB |
| RD_NUM_OF_FILES | Number of files to read |
| WT_FILE_SIZE | Size of a file to write in MB |
| WT_NUM_OF_FILES | Number of files to write |

Table 8: Enhanced DFSIO Parameters

For the Enhanced DFSIO benchmark, the file sizes (parameters RD_FILE_SIZE and WT_FILE_SIZE), which the workload should read and write, were fixed to 100MB. In the same time, the number of files (parameters RD_NUM_OF_FILES and RD_NUM_OF_FILES) were fixed to be 100, 200 and 500 to operate on a data set with data sizes of 10, 20 and 50 GB. The total data size is the product of multiplying the specific file size with the number of files to be read/written. Three experiments were executed as listed in Table 9.

| Data Size (GB) | RD_FILE_SIZE | RD_NUM_OF_FILES | WT_FILE_SIZE | WT_NUM_OF_FILES |
|---|---|---|---|---|
| 10 | 100 | 100 | 100 | 100 |
| 20 | 100 | 200 | 100 | 200 |
| 50 | 100 | 500 | 100 | 500 |

Table 9: Enhanced DFSIO Experiments

### 5.2.2. Results and Evaluation

The first subsection compares the performance between the Standard and Data-Computer cluster configurations. In the second subsection we compare and evaluate how increasing the size of pro-cessing data changes the performance for each cluster configuration. In both subsections the *Enhanced DFSIO-read* and *Enhanced DFSIO-write* parts are presented and discussed separately.

#### 5.2.2.1. Comparing Different Cluster Configurations

Figure 10 depicts the normalized *Enhanced DFSIO-read* times, with Standard1 (3 WNs) achieving the best times for all test cases. Table 10 compares the cluster configurations utilizing the same number of VMs, whereas Table 11 compares configurations utilizing different number of VMs.

In the first case, Standard1 (3 WNs) performs up to 73% better than Data-Comp1 (2 CWNs & 1 DWN) because of the different data placement strategies. In Data-Comp1 the data is stored on a single data node and should be read in parallel by the two compute nodes, which is not the case in Standard1, where each node stores the data locally, avoiding any communication conflicts. In the second case, Standard2 (6 WNs) performs between 18-46% slower than Data-Comp3 (3 CWNs & 3 DWNs). Although, each node in Standard2 stores a local copy of the data, it seems that the resources allocated per VM are not sufficient to run both *TaskTracker* and *DataNode* services, which is not the case in Data-Comp3.

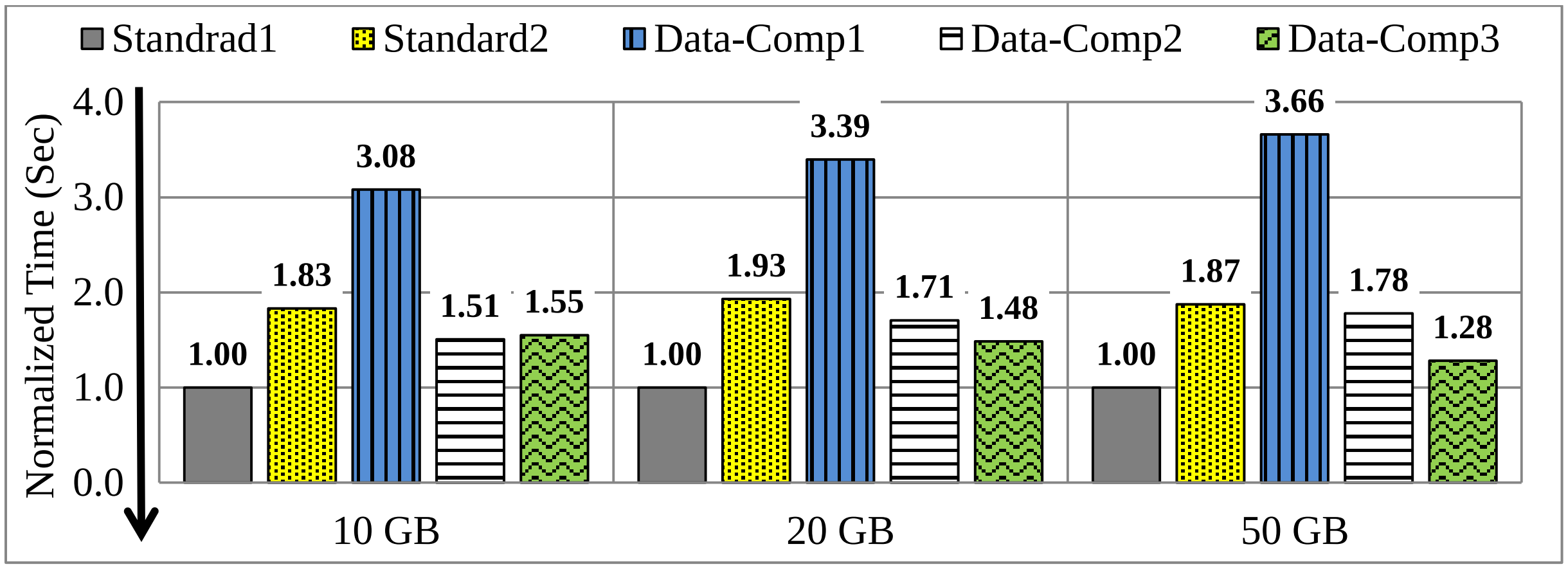

Figure 10: Normalized DFSIO Read Completion Times

| Equal Number of VMs | 3 VMs | 6 VMs |
|---|---|---|
| **Data Size (GB)** | **Diff. (%) Standard1/ Data-Comp1** | **Diff. (%) Standard2/ Data-Comp3** |
| 10 | +68 | -18 |
| 20 | +71 | -30 |
| 50 | +73 | -46 |

Table 10. DFSIO Read - *Equal Number of VMs*

| Different Number of VMs | 3 VMs / 4 VMs | 4 VMs / 6 VMs |
|---|---|---|
| **Data Size (GB)** | **Diff. (%) Data-Comp1/ Data-Comp2** | **Diff. (%) Data-Comp2/ Data-Comp3** |
| 10 | -104 | +3 |
| 20 | -99 | -15 |
| 50 | -106 | -39 |

Table 11. DFSIO Read - *Different Number of VMs*

In Table 11 we observe that Data-Comp2 (2 CWNs & 2 DWNs) completion times are *two times* faster than Data-Comp1 (2 CWNs & 1 DWN). On the other hand, Data-Comp3, which utilizes 3 data nodes, is up to 39% faster than Data-Comp2. This complies with our assumption that using more data nodes improves the read performance.

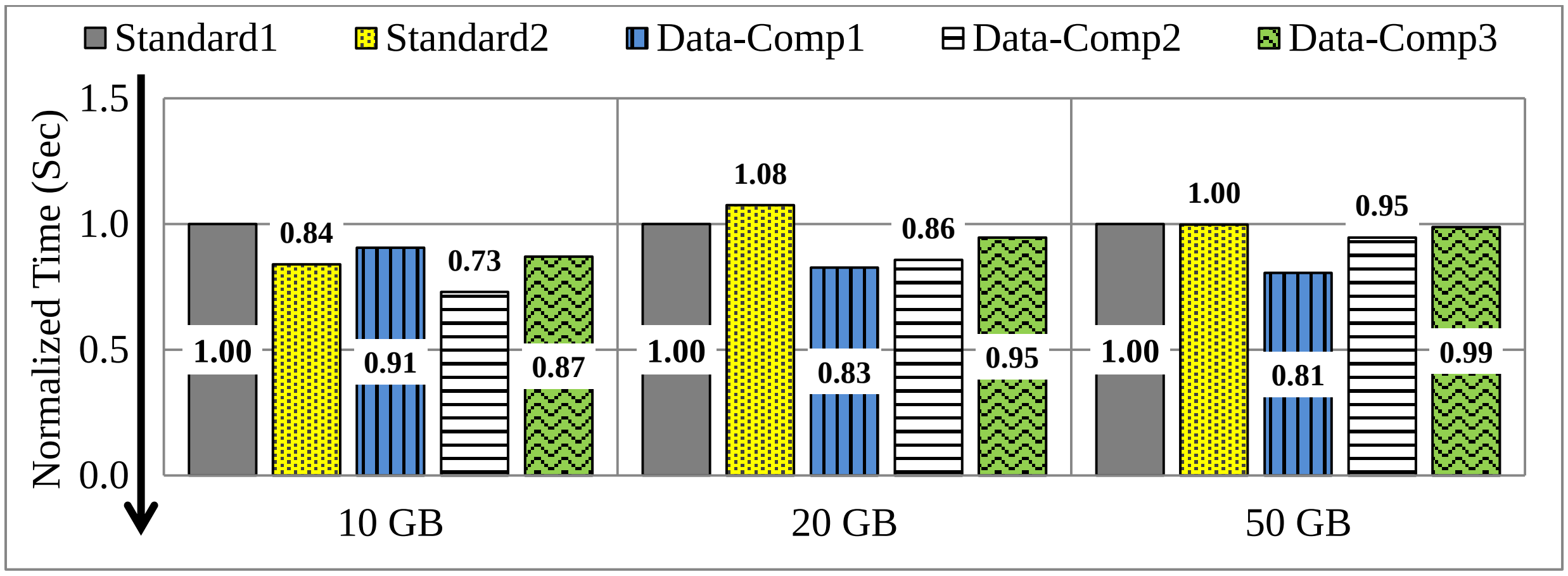

Figure 11: Normalized DFSIO Write Completion Times

| Equal Number of VMs | 3 VMs | 6 VMs |
|---|---|---|
| Data Size (GB) | Diff. (%) Standard1/ Data-Comp1 | Diff. (%) Standard2/ Data-Comp3 |
| 10 | -10 | +4 |
| 20 | -21 | -14 |
| 50 | -24 | -1 |

Table 12. DFSIO Write - *Equal* Number of VMs

| Different Number of VMs | 3 VMs / 6 VMs | 3 VMs / 6 VMs |
|---|---|---|
| Data Size (GB) | Diff. (%) Data-Comp1/ Data-Comp3 | Diff. (%) Standard1/ Data-Comp3 |
| 10 | -4 | -15 |
| 20 | +13 | -6 |
| 50 | +19 | -1 |

Table 13. DFSIO Write - *Different* Number of VMs

Figure 11 illustrates the *Enhanced DFSIO-write* [30] completion times for the five cluster configurations. Table 12 compares the cluster configurations utilizing the same number of VMs. In the first case, Standard1 (3 WNs) performs between 10-24% slower than Data-Comp1 (2 CWNs & 1 DWN). The reason for this is that Data-Comp1 utilizes only one data node and the HDFS pipeline writing process writes all three block copies locally on the node, which of course is against the fault tolerance practices in Hadoop. In a similar way, Data-Comp3 (3 CWNs & 3 DWNs) achieves between 1- 14% better times than Standard2 (6 WNs).
Table 13 compares cluster configurations with different number of VMs. Data-Comp1 (2 CWNs & 1 DWN) achieves up to 19% better times than Data-Comp3 (3 CWNs & 3 DWNs), because of the extra cost of writing to 3 data nodes (enough to guarantee the minimum data fault tolerance) instead of only one data node. Further observations show that although Data-Comp3 utilizes 6VMs, it achieves up to 15% better times than Standard1, which utilizes only 3VMs. However, this difference decreases from 15% to 1% with the growing data sizes and may completely vanish for larger data sets.

#### 5.2.2.2. Processing Different Data Sizes

Figure 12 shows the *Enhanced DFSIO-read* processing times (in seconds) for the different data sizes for the five tested cluster configurations. *The shorter times indicate better performance and respectively the longer times indicate worse performance.* Clearly, cluster configuration Standard1 (3 WNs) achieves the fastest times for all the three data sizes compared to the other configurations. This is also observed on Figure 13, which depicts the throughputs (MBs per second) for the five configurations, where configuration Standard1 (3 WNs) achieves the highest throughput between 143-147 MBs per second.

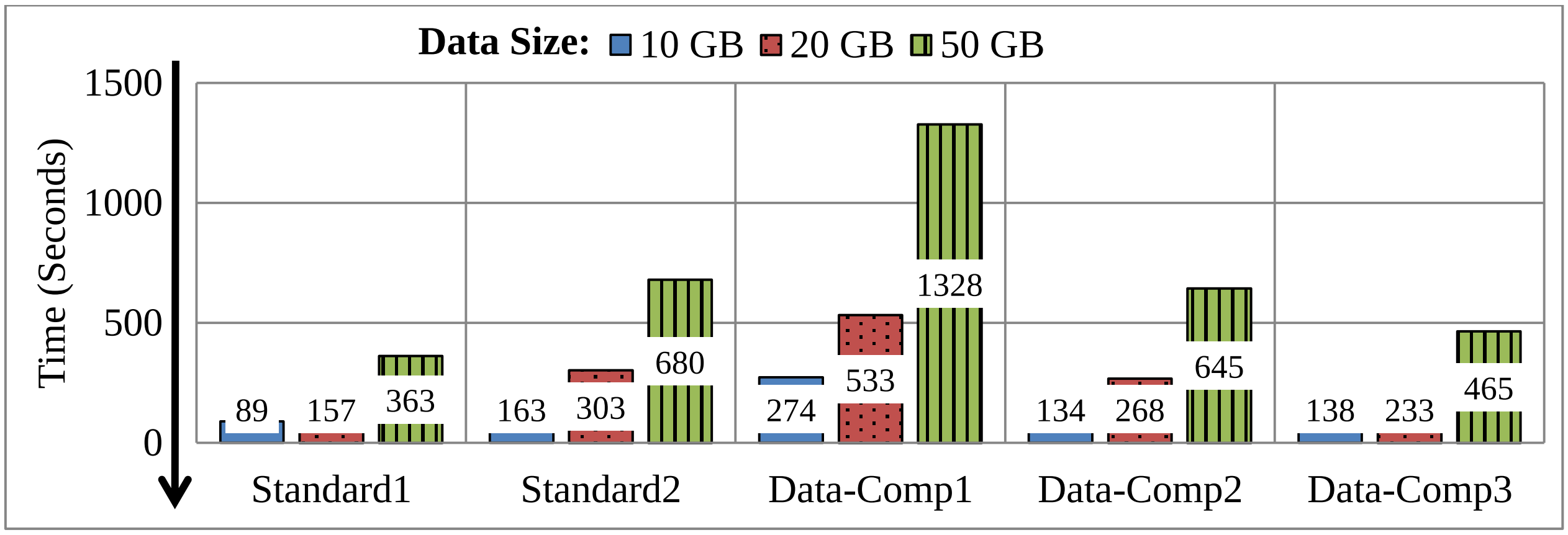

Figure 12: DFSIO Read Time (Seconds)

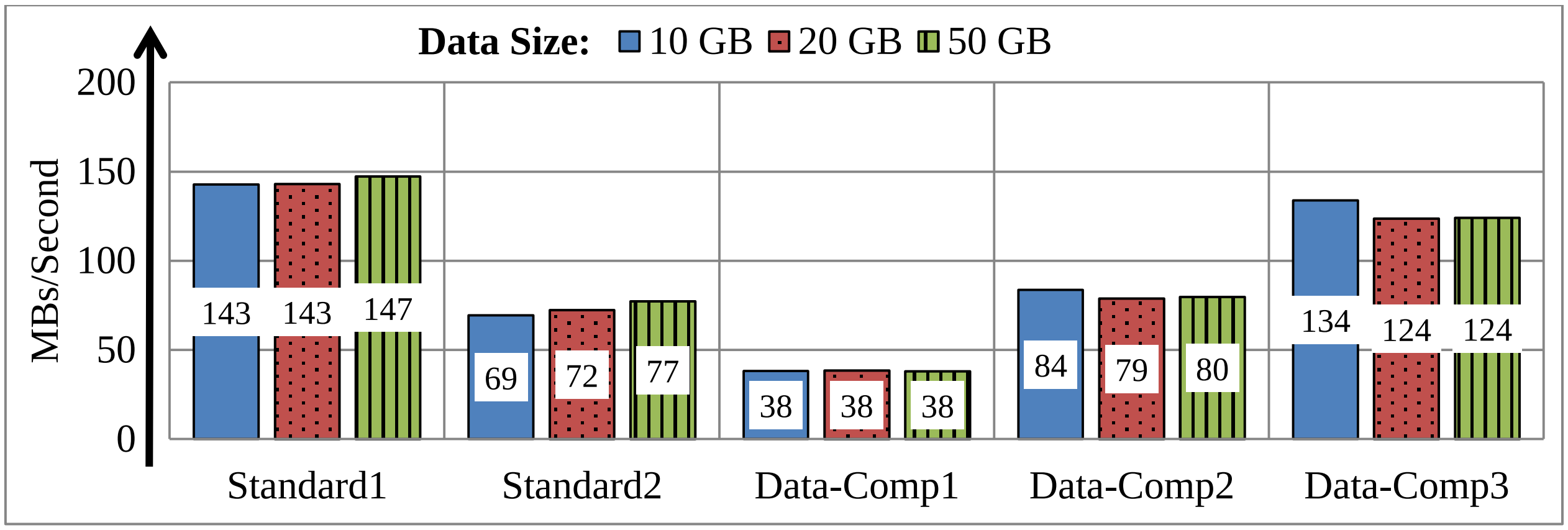

Figure 13: DFSIO Read Throughput (MBs per second)

The following Table 14 and Table 15 summarize the processing times for the tested Standard and Data-Compute cluster configurations. Additionally, there is a column "**Data Δ"** representing the data increase in percent compared to the baseline data size, which is 10GB. For example, the **Δ** between the baseline (10GB) and 20GB is +100% and respectively for 50GB is +400%. Also, there are multiple columns "**Time Δ**", one per cluster configuration, which indicates the time difference in percent compared to the time of Standard1 (3 WNs), which we use as the *baseline configuration*. For example, comparing the times for processing 10GB of the baseline Standard1 (3 WNs) with Standard2 (6 WNs) configuration results in 83.15% time difference, as shown in Table 14. This means that Standard2 needs 83.15% more time compared to the baseline Standard1 to read the 10GB data. On the contrary, the *negative time differences* indicate faster completion times in comparison to the baseline.

| **Data Size (GB)** | **Data Δ (%)** | **Read Time (Sec)** | | |
|---|---|---|---|---|
| | | **Standard1 Baseline** | **Standard2** | **Time Δ (%)** |
| 10 | baseline | 89 | 163 | 83.15 |
| 20 | +100 | 157 | 303 | 92.99 |
| 50 | +400 | 363 | 680 | 87.33 |

Table 14: DFSIO Standard Cluster Read Results

| Data Size (GB) | Data Δ (%) | Read Time (Sec) | | | | | |
|---|---|---|---|---|---|---|---|
| | | Data-Comp1 | Time Δ (%) | Data-Comp2 | Time Δ (%) | Data-Comp3 | Time Δ (%) |
| 10 | baseline | 274 | 207.87 | 134 | 50.56 | 138 | 55.06 |
| 20 | +100 | 533 | 239.49 | 268 | 70.70 | 233 | 48.41 |
| 50 | +400 | 1328 | 265.84 | 645 | 77.69 | 465 | 28.10 |

Table 15: DFSIO Data-Compute Cluster Read Results

Figure 14 illustrate the time differences in percent of all the tested cluster configurations normalized with respect to Standard1 (3 WNs) as baseline. We observe that all time differences are positive, which means that they perform slower than the baseline configuration. Data-Comp3 (3 CWNs & 3 DWNs) configuration has the smallest time difference ranging between 28.10% and 55.06%, whereas the worst performing configuration is Data-Comp1 (2 CWNs & 1 DWN) with time differences between 207.87% and 265.84%.

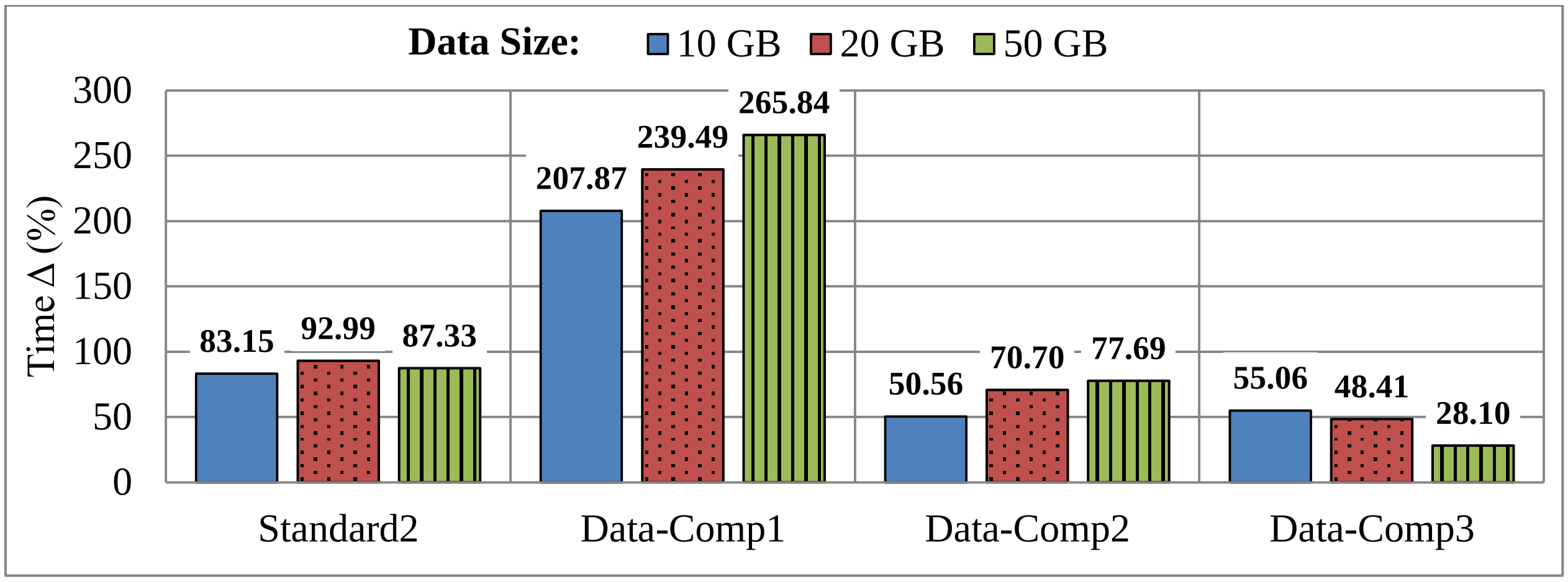

Figure 14: DFSIO Read Time Difference between Standard1 (Baseline) and all Other Configurations in %

Figure 15 illustrates how the different cluster configurations scale with the increasing data sets normalized to the baseline configuration. We observe that all configurations scale almost linearly with the increase of the data sizes. Observing the graph we can clearly distinguish that Standard1 (3 WNs) is the fastest configuration as its line lies much lower than all other configurations. On the contrary, Data-Comp1 (2 CWNs & 1 DWN) is the slowest configuration as its data points are the highest one for all three data sizes.

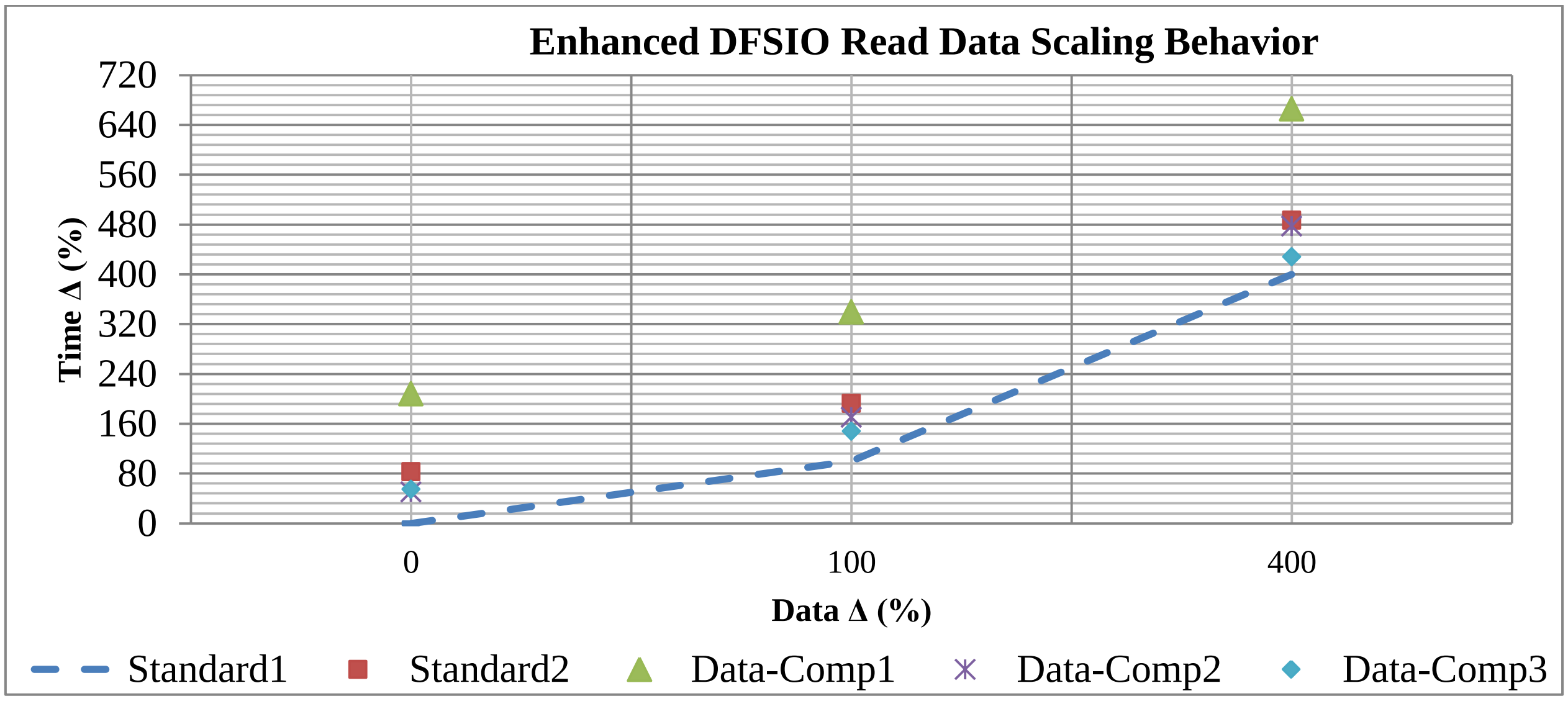

Figure 15: DFSIO Read Data Behavior of all Cluster Configurations normalized to Standrad1

Figure 16 shows the *Enhanced DFSIO-write* processing times (in seconds) for the different data sizes for the five tested cluster configurations. *The shorter times indicate better performance and respectively the longer times indicate worse performance*. If we look more closely at Figure 16, we can identify that cluster configuration Data-Comp1 (2 CWNs & 1 DWN) achieves the fastest times for 20GB and 50GB data sizes. This can be also observed on Figure 17, which depicts the throughputs (MBs per second) for the five configurations, where configuration Data-Comp1 (2 CWNs & 1 DWN) achieves the highest throughput around 67-68 MBs per second.

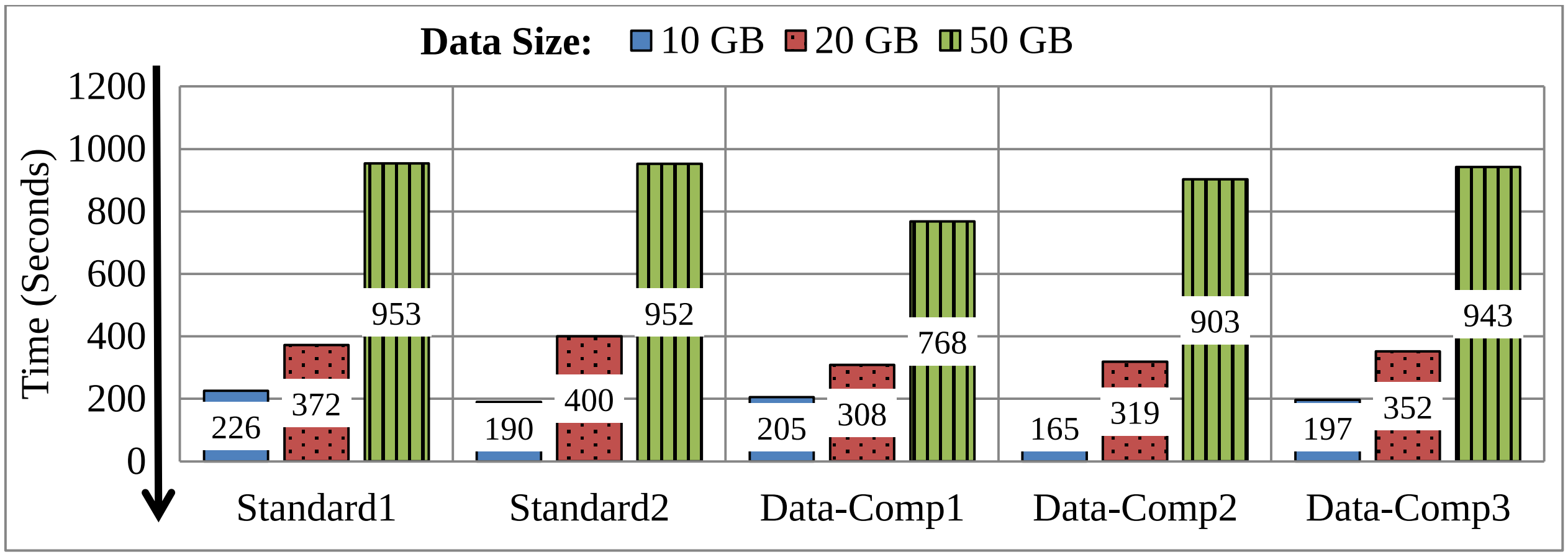

Figure 16: DFSIO Write Time (Seconds)

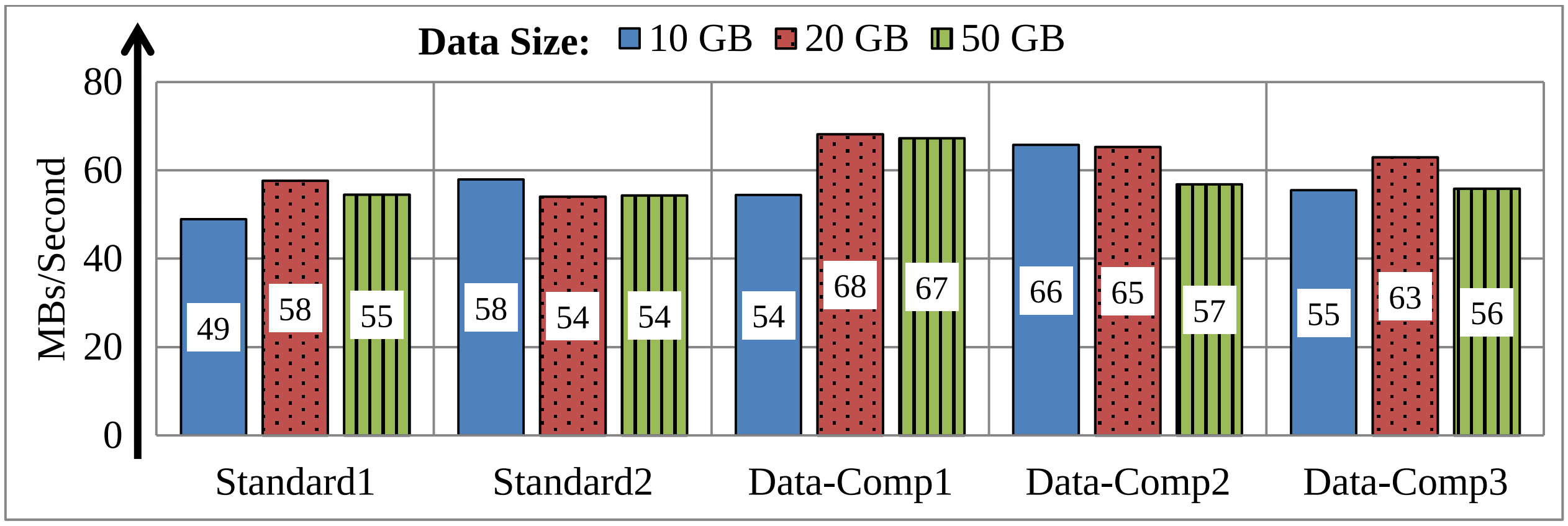

Figure 17: DFSIO Write Throughput (MBs per second)

The following Table 16 and Table 17 summarize the processing times for the tested Standard and Data-Compute cluster configurations. Additionally, there is a column "**Data Δ**" representing the data increase in percent compared to the baseline data size, which is 10GB. For example, the **Δ** between the baseline (10GB) and 20GB is +100% and respectively for 50GB is +400%. Also, there are multiple columns "**Time Δ**", one per cluster configuration, which indicates the time difference in percent compared to the time of Standard1 (3 WNs), which we use as the *baseline configuration*. For example, comparing the times for processing 10GB of the baseline Standard1 (3 WNs) with Standard2 (6 WNs) configuration results in -15.93% time difference. This means that Standard2 finish for 15.93% less time compared to the baseline Standard1. On the contrary, the *positive time differences* indicate slower completion times in comparison to the baseline.

| **Data Size (GB)** | **Data Δ (%)** | **Write Time (Sec)** | | |
|---|---|---|---|---|
| | | **Standard1 Baseline** | **Standard2** | **Time Δ (%)** |
| 10 | baseline | 226 | 190 | -15.93 |
| 20 | +100 | 372 | 400 | +7.53 |
| 50 | +400 | 953 | 952 | -0.10 |

Table 16: DFSIO Standard Cluster Write Results

| **Data Size (GB)** | **Data Δ (%)** | **Write Time (Sec)** | | | | | |
|---|---|---|---|---|---|---|---|
| | | **Data-Comp1** | **Time Δ (%)** | **Data-Comp2** | **Time Δ (%)** | **Data-Comp3** | **Time Δ (%)** |
| 10 | baseline | 205 | -9.29 | 165 | -26.99 | 197 | -12.83 |
| 20 | +100 | 308 | -17.20 | 319 | -14.25 | 352 | -5.38 |
| 50 | +400 | 768 | -19.41 | 903 | -5.25 | 943 | -1.05 |

Table 17: DFSIO Data-Compute Cluster Write Results

Figure 18 depicts the time differences in percent of all the tested cluster configurations normalized with respect to Standard1 (3 WNs) as baseline. We observe that all time differences except for the 20GB experiment with Standard2 configuration are negative, which means that they perform faster than the baseline configuration. Data-Comp1 (2 CWNs & 1 DWN) and Data-Comp2 (2 CWNs & 2 DWN) achieve the highest time differences ranging between -5.25% and -26.99% making them the best performing cluster configurations.

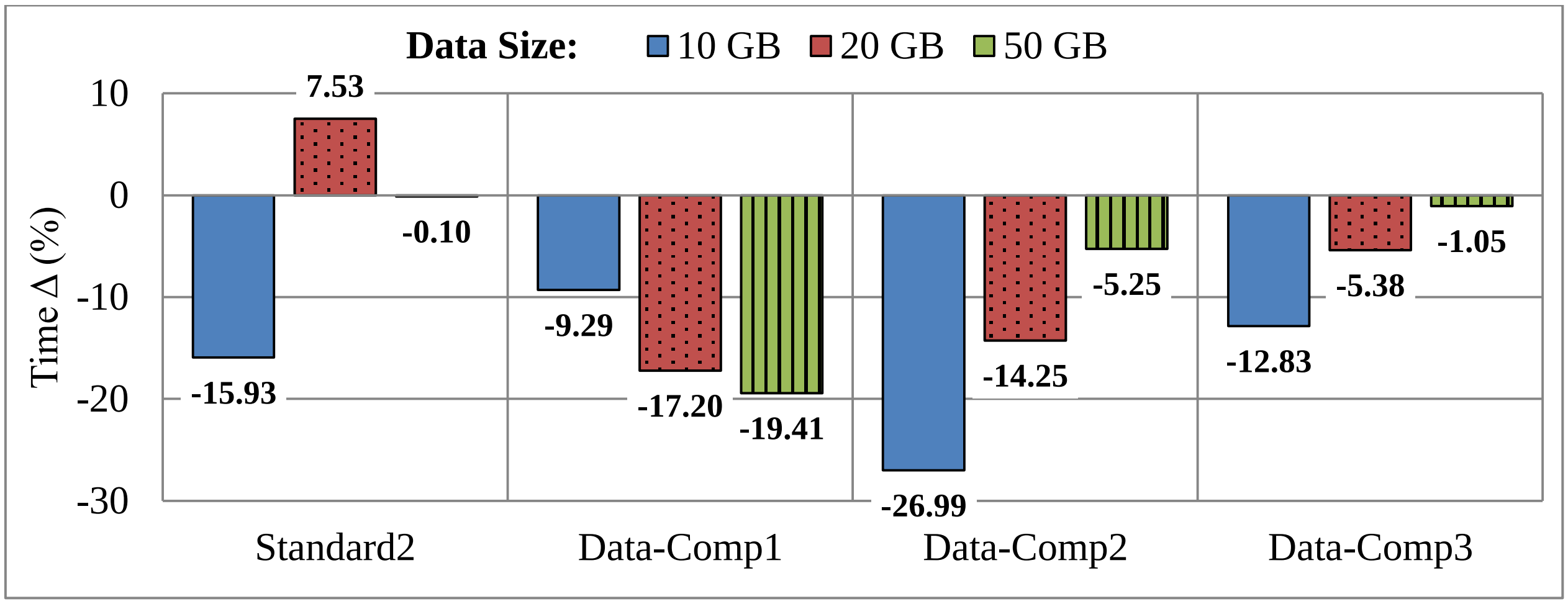

Figure 18: DFSIO Write Time Δ (%) between Standard1 (Baseline) and all other configurations in %

Figure 19 illustrates how the different cluster configurations scale with the increasing data sets normalized with respect to the baseline configuration. In this case, we observe that all configurations scale almost linearly with the increase of the data sizes, although the time differences are varying. Looking closely at the graphic we can distinguish that Standard1 (3 WNs) is the slowest configuration as its line lies slightly higher than most of the other configurations. On the contrary, Data-Comp1 (2 CWNs & 1 DWN) and Data-Comp2 (2 CWNs & 2 DWN) are the fastest configuration as their data points are the lowest one for all three data sizes.

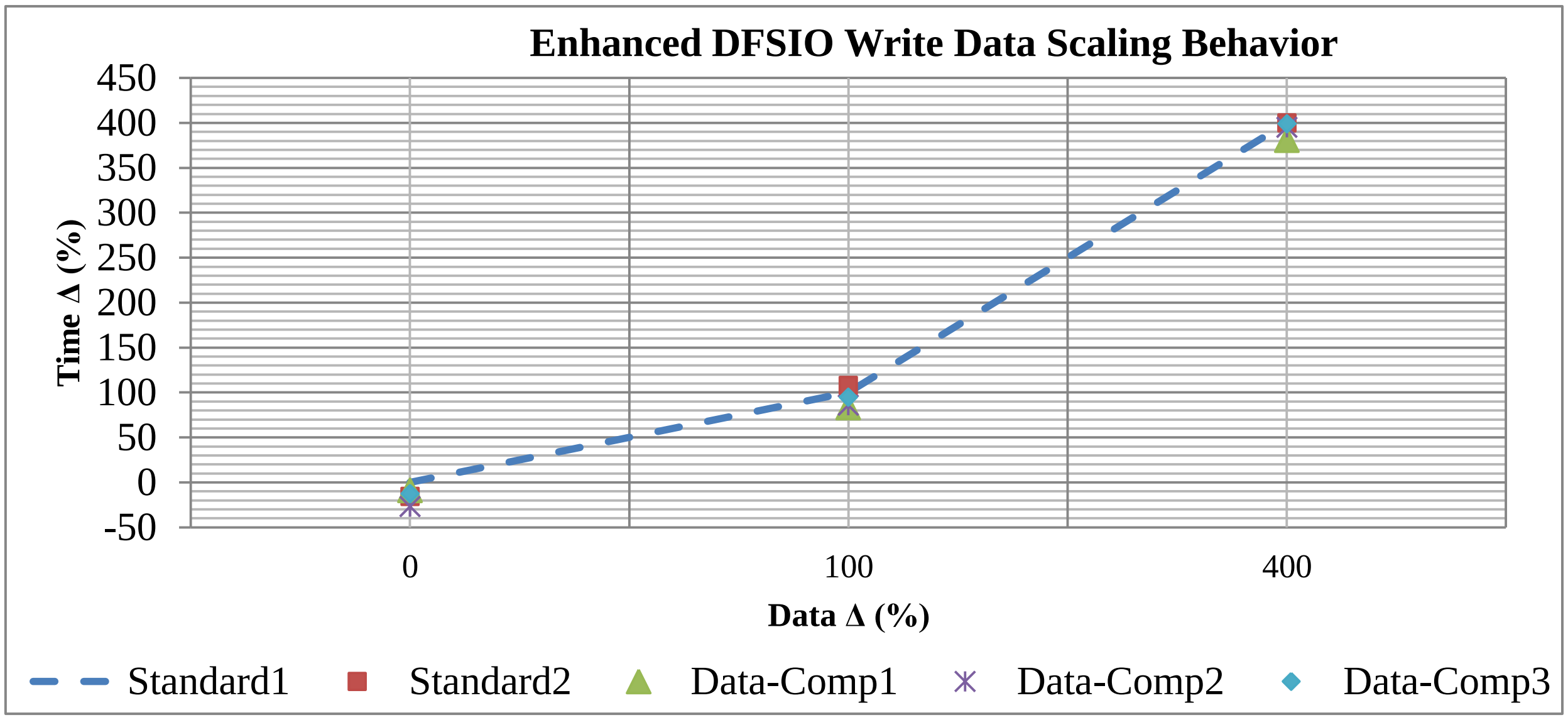

Figure 19: DFSIO Write Data Behavior of all Cluster Configurations normalized to Standrad1

Overall, our experiments showed that the *Enhanced DFSIO-read* and *Enhanced DFSIO-write* workloads scale nearly linear with the increase of the data size. We observed that Standard1 (3 WNs) achieves the slowest times for *DFSIO-read*, whereas Data-Comp1 (2 CWNs & 1 DWN) achieves the fastest *DFSIO-write* times.

## 6. Lessons Learned

Our experiments showed:

- *Compute-intensive* (i.e. CPU bound WordCount) workloads are more suitable for Standard Hadoop clusters. However, we also observed that adding more compute nodes to a Data-Compute cluster improves the performance of CPU bound applications (see Table 4).
- *Read-intensive* (i.e. read I/O bound DFSIO) workloads perform best (Standard1) when hosted on a Standard Hadoop cluster (see Table 10). However, adding more data nodes to a Data-Compute Hadoop cluster improved up to 39% the reading speed (e.g. Data-Comp2/Data-Comp3, see Table 11).
- *Write-intensive* (i.e. write I/O bound DFSIO) workloads were up to 15% faster (e.g. Standard2/Data-Comp3 and Standard1/Data-Comp3, see Table 12 and Table 13) on a Data-Compute Hadoop cluster in comparison to a Standard Hadoop cluster. Also our experiments showed that using less data nodes results in better write performance (e.g. Data-Comp1/Data-Comp3) on a Data-Compute Hadoop cluster, reducing the overhead of data transfer.

In addition it must be noted that Data-Compute cluster configurations are more advantageous in respect to node elasticity [33]. Therefore, the overhead for read- or compute-intensive workloads might be acceptable.

During the benchmarking process, we identified three important factors which should be taken into account when configuring a virtualized Hadoop cluster:

- *Choosing the "right" cluster type (Standard or Data-Compute Hadoop cluster) that provides the best performance for the hosted Big Data workload is not a straightforward process.* It requires very precise knowledge about the workload type, i.e. whether it is CPU intensive, I/O intensive or mixed, as indicated in Section 4.
- *Determining the number of nodes for each node type (compute and data nodes) in a Data-Compute cluster is crucial for the performance and depends on the specific workload characteristics.* The extra network overhead, caused by intensive data transfer between data and compute worker nodes, should be carefully considered, as also reported by Ye et al. [32].
- *The overall number of virtual nodes running in a cluster configuration has direct influence on the workload performance*, this is also confirmed by [31],[6],[32]. Therefore, it is crucial to choose the optimal number of virtual nodes in a cluster, as each additional VM causes an extra overhead to the hypervisor. At the same time, we observed cases, e.g. *Standard1/Standard2* and *Data-Comp1/Data-Comp2*, where clusters consisting of more VMs utilized better the underlying hardware resources.

# Appendix

The Serengeti Server JSON file defines the Data-Comp1 cluster configuration.

```
{
"nodeGroups" : [
{
"name" : "DataMaster",
"roles" : [
"hadoop_namenode"
],
"instanceNum" : 1,
"instanceType" : "SMALL",
"storage" : {
"type" : "shared",
"shares" : "NORMAL",
"sizeGB" : 25
},
"cpuNum" : 1,
"memCapacityMB" : 3748,
"swapRatio" : 1.0,
"haFlag" : "on",
"configuration" : {
"hadoop" : {
}
}
},
{
"name" : "ComputeMaster",
"roles" : [
"hadoop_jobtracker"
],
"instanceNum" : 1,
"instanceType" : "SMALL",
"storage" : {
"type" : "shared",
"shares" : "NORMAL",
"sizeGB" : 25
},
"cpuNum" : 1,
"memCapacityMB" : 3748,
"swapRatio" : 1.0,
"haFlag" : "on",
"configuration" : {
"hadoop" : {
}
}
},
{
"name" : "ComputeWorker",
"roles" : [
"hadoop_tasktracker"
],
"instanceNum" : 2,
"instanceType" : "SMALL",
"storage" : {
"type" : "shared",
"shares" : "NORMAL",
"sizeGB" : 50
},
"cpuNum" : 5,
"memCapacityMB" : 4608,
"swapRatio" : 1.0,
"haFlag" : "off",
"configuration" : {
"hadoop" : {
}
}
},
{
"name" : "DataWorker",
"roles" : [
"hadoop_datanode"
],
"instanceNum" : 1,
"instanceType" : "SMALL",
"storage" : {
"type" : "shared",
"shares" : "NORMAL",
"sizeGB" : 200
},
"cpuNum" : 2,
"memCapacityMB" : 4608,
"swapRatio" : 1.0,
"haFlag" : "off",
"configuration" : {
"hadoop" : {
}
}
},
{
"name" : "Client",
"roles" : [
"hadoop_client",
"pig",
"hive",
"hive_server"
],
"instanceNum" : 1,
"instanceType" : "SMALL",
"storage" : {
"type" : "shared",
"shares" : "NORMAL",
"sizeGB" : 50
},
"cpuNum" : 1,
"memCapacityMB" : 3748,
"swapRatio" : 1.0,
"haFlag" : "off",
"configuration" : {
"hadoop" : {
}
}
}
],
"configuration" : {
},
"specFile" : false
}
```

## Acknowledgements

This research is supported by the Big Data Lab at the Chair for Databases and Information Systems (DBIS) of the Goethe University Frankfurt. We would like to thank Alejandro Buchmann of Technical University Darmstadt, Nikolaos Korfiatis and Jeffrey Buell of VMware for their helpful comments and support.